\documentclass[12pt]{article}
\usepackage{epsfig,amsfonts,amssymb}
\usepackage{hyperref}
\pdfoutput=1

\usepackage{cite}
\usepackage{cite}

\usepackage{comment}

\def\ZZZ{{\hbox{ Z\kern-1.6mm Z}}}
\def\RRR{{\hbox{ R\kern-2.4mm R}}}
\def\CCC{{\hbox{ C\kern-2.0mm C}}}
\def\zzz{{\hbox{z\kern-1mm z}}}

\newcommand{\qeq}{{\hbox{=\kern-2.3mm ? \kern.5mm }}}
\renewcommand{\qeq}{=}

\newcommand{\DD}{{\cal D}}

\newcommand{\NN}{{\cal N}}

\newcommand{\be}{\begin{equation}}
\newcommand{\ee}{\end{equation}}
\newcommand{\ben}{\begin{eqnarray}\displaystyle}
\newcommand{\een}{\end{eqnarray}}

\newcommand{\refb}[1]{(\ref{#1})}
\newcommand{\p}{\partial}
\newcommand{\sectiono}[1]{\section{#1}\setcounter{equation}{0}}

\def\one{{\hbox{ 1\kern-.8mm l}}}
\def\zero{{\hbox{ 0\kern-1.5mm 0}}}

\newcommand{\bea}[1]{\begin{eqnarray}\label{#1} }
\newcommand{\eea}{\end{eqnarray}}

\newcommand{\eqref}{\refb}

\newcommand{\q}{e}

\usepackage{bm}
\usepackage[table]{xcolor}

\begin{document}

\baselineskip 24pt

\begin{center}

{\Large \bf Logarithmic Correction to BPS Black Hole Entropy from 
Supersymmetric Index at Finite Temperature}

\end{center}

\vskip .6cm
\medskip

\vspace*{4.0ex}

\baselineskip=18pt

\centerline{\large \rm A.H.~Anupam, P.V.~Athira,  Chandramouli Chowdhury  and Ashoke Sen}

\vspace*{4.0ex}

\centerline{\large \it International Centre for Theoretical Sciences - TIFR 
}
\centerline{\large \it  Bengaluru - 560089, India}


\vspace*{1.0ex}
\centerline{\small E-mail: anupam.ah, athira.pv,
chandramouli.chowdhury,
 ashoke.sen \ @icts.res.in}

\vspace*{5.0ex}

\centerline{\bf Abstract} \bigskip

It has been argued by Iliesiu, Kologlu and Turiaci in arXiv:2107.09062
that one can compute the supersymmetric
index of black holes using  black hole geometry carrying finite temperature
but a specific complex angular velocity. We follow their prescription to compute the
logarithmic correction to the entropy of BPS states in four dimensions, 
defined as the log of 
the index of supersymmetric black holes, and find perfect agreement
with the previous results for the same quantity computed using the near horizon $AdS_2\times
S^2$ geometry of zero temperature black holes.
Besides giving an independent computation of supersymmetric black hole
entropy, this analysis also provides a test of the procedure used previously for computing
logarithmic corrections to Schwarzschild and other non-extremal black hole entropy.

\vfill \eject

\tableofcontents

\sectiono{Introduction}

Logarithmic corrections to extremal, supersymmetric black hole entropy has provided a
stringent consistency test of string theory. In particular, logarithmic corrections to 
supersymmetric black  hole entropy in $\NN=4$ and $\NN=8$ supersymmetric
string theories in
four dimensions,
computed from gravitational path integral over the near horizon $AdS_2\times S^2$
geometry\cite{1005.3044,1106.0080}, 
agrees with the
result of exact counting in the corresponding string 
theories\cite{9607026,9903163,0505094,0506249,0510147,0605210,
0607155,0609109,1008.3801}.\footnote{Throughout this paper we
shall use the words entropy and log of the index interchangeably since at a generic point
in the moduli space of the theory, we expect all the unprotected states to be lifted and only
the states protected by the index should form the degenerate ground state.
}
Similar agreement
also holds for supersymmetric black holes in higher dimensional string 
theories\cite{1109.3706}.
A summary
of some of the results can be found in table 1 of \cite{1205.0971}.
This analysis has also been extended to 
exponentially suppressed contribution to the indices
and for twisted indices\cite{1402.2441,1404.6363} and to black holes in
asymptotically anti de Sitter spaces\cite{1711.01076,2008.03239,2106.09730 }.

In a relatively recent paper, Iliesiu, Kologlu and Turiaci have suggested that we can compute
the index of supersymmetric black holes in flat space-time
by working in a finite temperature black hole 
geometry\cite{2107.09062}. (Some earlier work along the same line
on black holes in AdS space-time can be found in \cite{1810.11442}
and on near extremal black holes
can be found in  \cite{Heydeman:2020hhw}.)
Typically at finite temperature the Euclidean path integral around
a black hole geometry is expected to compute the partition function in the grand canonical
ensemble. In particular the angular velocity $\vec\Omega$ of the black hole
provides the
chemical potential dual to the angular momentum $\vec J$ carried by the black hole, and
the asymptotic values of the time component $\mu_i$ of the $i$-th gauge field
 provide the chemical potential dual to the electric 
 gauge charge. 
 Therefore the gravitational path
 integral computes
 \be\label{e1}
 Tr_{\vec P}\left(e^{-\beta E -\beta \mu_i Q_i - \beta \,\vec\Omega.\vec J}\right)\, ,
 \ee
 where the trace is taken over all states at fixed magnetic charges $\vec P$. 
 The components of $\vec P$ are fixed
 by the fluxes of the magnetic fields at infinity. In contrast, a supersymmetric
 black hole at zero temperature 
 develops a near horizon $AdS_2\times S^2$ geometry and the Euclidean path integral over
 this geometry gives the degeneracy / index of the black hole in the microcanonical 
 ensemble\cite{0809.3304} where the electric and the magnetic charges, as well as
 the angular momentum, are fixed.
 
 The main idea behind \cite{2107.09062} is that if we work at fixed $\beta, \mu_i$ and set
 $\beta\, \vec\Omega=(0,0,-2\pi i)$ in \refb{e1}, 
 then the part inside the trace involving the angular momentum
 becomes $e^{2\pi i J_3}=(-1)^F$ since $J_3$ is integer (integer + half) for bosons
 (fermions). Hence the path integral will now measure the
 supersymmetric index.\footnote{As we shall see, for the special case when the magnitude of
 the vector
 $i\beta\, \vec\Omega$ is $2\pi$ and hence $e^{-\beta\,\vec\Omega.\vec J}$ is independent
 of the direction of $\vec\Omega$,
 there is a family of supersymmetric saddles
 related by rotation of the vector $\vec\Omega$, and we have to integrate over this family
 to get the correct result for the entropy. \label{f2}} 
 Since supersymmetric index receives contribution only from 
 BPS states that carry lowest possible energy for a given charge,
 the index computation behaves
 as if we are at zero temperature
 even if we keep $\beta$ finite. 
 Indeed, it was shown in \cite{2107.09062} that requiring the solution to have
 $\beta\, \vec\Omega=(0,0,-2\pi i)$ sets the black hole mass $M$ to be its BPS
 bound.
 Therefore even by working with a finite temperature black hole we
 can study the contribution to the supersymmetric index from BPS black holes. 
 However we still have a sum over electric charges in \refb{e1}. 
 In order to recover
 the index for a given electric charge from \refb{e1} we need to perform a Laplace
 transform. In the large charge limit this can be performed using saddle point technique.
 It was shown in \cite{1205.0971} that in four space-time dimensions the Laplace transform
 does not generate any logarithmic correction to the log of the index.  Therefore for
 computing logarithmic correction to the index we can continue to use the
 Euclidean gravitational path integral describing \refb{e1}, and fix $\mu_i$ in terms
 of $Q_i$'s using the classical solution describing the black hole. 
 
 There is, however, an additional subtlety that we need to address.
 The Witten index of a supersymmetric
 black hole defined in \refb{e1} 
 is known to vanish since the black hole always breaks some of the
  supersymmetries and quantization of the associated goldstino modes produces Bose-Fermi
  degenerate pair of states. In particular if there are $2n$ broken supersymmetries then
  there are $2n$ such goldstino fermion zero modes. The remedy is to insert 
 $n$ factors of $2J_3$ carried
  by the state into the trace so that the cancellation is prevented for states that break
  only $2n$ supersymmetries, but continue to hold for states that break more than $2n$ 
  supersymmetries\cite{9611205,9708062}. 
  This small modification of the prescription of \cite{2107.09062} will be needed
  for our analysis.

Our goal will be to compute the logarithmic correction to the supersymmetric
black hole entropy following
the prescription given above and compare the result with the zero temperature 
computation based on the near horizon $AdS_2\times S^2$ 
geometry\cite{1005.3044,1106.0080,1108.3842,1109.0444,1404.1379, Iliesiu:2022onk}. 
We find perfect agreement between the two computations.
In the following we
shall refer to the full black hole geometry at finite temperature as the \emph{
finite temperature
geometry} and the near horizon $AdS_2\times S^2$ geometry of a zero temperature black
hole as the \emph{zero temperature geometry}.

The rest of the paper is organized as follows. In section \ref{sreview} we describe the
Kerr-Newman solution in the scaled variable that makes it easier to extract the
dependence of various quantities on the overall scale of the charge carried by the black
hole. We also review how to normalize the integration measure over various fields in the
gravitational path integral. In section \ref{sb} we discuss a pair of rotational zero modes
of four dimensional Euclidean black holes that arise specifically for the special choice of
$\beta\, \vec\Omega=(0,0,-2\pi i)$. These are related to the freedom of rotating $\vec\Omega$
as mentioned in footnote \ref{f2}. 
In section \ref{slog} we describe the computation of logarithmic 
corrections to the entropy of supersymmetric black holes in $\NN\ge 2$ string theories.
We conclude in section \ref{sconc} with some comments on our result. Throughout this paper
we shall work in the $\hbar=c=G=1$ units.

\def\q{\theta}
\def\w{\wedge}

\sectiono{Review} \label{sreview}

In this section we shall review some of the background material needed for our analysis.

\subsection{The Kerr-Newman solution in scaled variables} \label{sa}

We begin by writing down the Kerr-Newman solution:
\ben\label{knboyer}
ds^2 &=& - {r^2 + a_0^2 \cos^2\theta - 2M r + Q^2\over 
r^2+ a_0^2 \cos^2\theta} \, dt^2 + \frac{r^2+ a_0^2 \cos^2\theta}{r^2 + a_0^2 - 2 M r + Q^2} dr^2 + (r^2+ a_0^2 \cos^2\theta) d\theta^2 \nonumber\\
&&+\frac{(r^2+ a_0^2 \cos^2\theta) (r^2 + a_0^2) + (2 M r - Q^2) a_0^2 \sin^2 \theta}{r^2+ a_0^2 \cos^2\theta} \sin^2\theta d\phi^2 \nonumber \\
&&+\frac{2(Q^2 - 2 M r)a_0}{r^2+ a_0^2 \cos^2\theta} \sin^2 \theta dt d\phi \, ,\nonumber \\
A_\mu dx^\mu &=& -{Q \, r\over r^2 + a_0^2 \cos^2 \theta} \left[ dt - a_0 \sin^2\theta
d\phi \right] +\, {Q \, r_+\over r_+^2 +a_0 ^2}  \, d t \, .
\een
In this geometry the outer and the inner horizons are located at
\be
r_\pm = M \pm \sqrt{M^2-Q^2-a_0^2}\, .
\ee
The inverse temperature $\beta$, the angular velocity $\Omega_3$
at the horizon and the chemical potential $\mu$ are given by:
\be\label{beta}
\beta = 4\pi {(r_+^2 + a_0^2)\,  r_+ \over r_+^2 - a_0^2 - Q^2}, \qquad
\beta\, \Omega_3 = {a_0\over r_+^2 + a_0^2} \beta= 4\pi {a_0  r_+ \over r_+^2 - a_0^2 - Q^2},
\qquad \mu=  {Q\, r_+\over r_+^2 + a_0^2}\, .
\ee
We shall see later that the constant additive term in  the expression for
$A_\mu dx^\mu$ in \refb{knboyer} is needed to get a
non-singular gauge field configuration at the horizon of the Euclidean black hole. For now we
note that the asymptotic value of $A_t$ determines the chemical potential $\mu$.

We shall consider an Euclidean version of this black hole and work with scaled variables
and coordinates defined as follows:
\be\label{echoice}
m= {M\over a}, \quad q= {Q\over a}, \quad b = i \, {a_0\over a},  \quad \rho = {r\over a}, \quad 
\tau
= i\, {t\over a}, \quad \rho_\pm={r_\pm\over a}, \quad \gamma={\beta\over a}, \quad
\vec\omega ={a\, \vec\Omega}\, ,
\ee
where $a$ is an arbitrary parameter. 
In this case the solution takes the
form:
\ben\label{eknscaled}
ds^2 &=& a^2\bigg[ {\rho^2 - b^2 \cos^2\theta - 2m \rho + q^2\over 
\rho^2- b^2 \cos^2\theta} \, d\tau^2 + \frac{\rho^2-b^2 \cos^2\theta}{\rho^2 - b^2 - 2 m
\rho + q^2} d\rho^2 + (\rho^2-b^2 \cos^2\theta) d\theta^2 \nonumber\\
&&+\frac{(\rho^2- b^2 \cos^2\theta) (\rho^2 -b^2) - (2 m\rho - q^2) b^2 \sin^2 \theta}{\rho^2-
b^2 \cos^2\theta} \sin^2\theta d\phi^2 \nonumber \\
&&-\frac{2(q^2 - 2 m\rho)b}{\rho^2-b^2 \cos^2\theta} \sin^2 \theta d\tau d\phi\bigg]\, ,
\nonumber \\
A_\mu dx^\mu &=& i\, a\, {q \, \rho \over \rho^2 -b^2 \cos^2 \theta} \left[ d\tau - b
\sin^2\theta\, d\phi \right] - i\, a\, {q \, \rho_+\over \rho_+^2 - b^2}  \, d\tau\, ,
\een
and we have
\be\label{esmall1}
\gamma = 4\pi {(\rho_+^2 - b^2)\,  \rho_+ \over \rho_+^2 +b^2 - q^2}, \qquad
\gamma\, \omega_3=\beta\, \Omega_3 =  -4\pi  i\, {b  \rho_+ \over \rho_+^2 +b^2 - q^2},
\qquad \mu={q\,  \rho_+ \over \rho_+^2 -b^2}\, ,
\ee
with
\be\label{esmall2}
\rho_\pm = m \pm \sqrt{m^2-q^2+b^2}\, .
\ee
The horizon corresponds to the surface $\rho=\rho_+$. 
To check the regularity at the horizon, we write the metric as
\ben
&&\hskip -.3in ds^2 = a^2\bigg[  \frac{\rho^2-b^2 \cos^2\theta}{\rho^2 - b^2 - 2 m
\rho + q^2} d\rho^2 + (\rho^2-b^2 \cos^2\theta) d\theta^2 \nonumber\\
&& +\,  {\rho^2 - b^2 - 2m\rho+q^2\over \rho^2-b^2\cos^2\theta}
\left(d\tau -b\sin^2\theta d\phi\right)^2 
+ {\sin^2\theta\over  \rho^2-b^2\cos^2\theta} \left\{ \left(\rho^2-b^2\right) d\phi 
+b d\tau\right\}^2
\bigg]\, .
\een
At $\rho=\rho_+$ the coefficient of $d\rho^2$ term blows up while the coefficient of the
$\left(d\tau -b\sin^2\theta d\phi\right)^2$ term vanishes. We now define
\be
\tilde \rho={\sqrt{\rho-\rho_+}}, \qquad \tilde \phi = \phi + {b\over \rho_+^2 - b^2}\tau\, ,
\ee
and express the metric and the gauge fields near the horizon as
\ben\label{escaledmetric}
&&\hskip -.3in ds^2 \simeq a^2\Bigg[  2\,  \frac{\rho_+^2-b^2 \cos^2\theta}{\sqrt{m^2-q^2+b^2}} \left\{d\tilde\rho^2 
+  {m^2-q^2+b^2\over (\rho_+^2-b^2)^2} \tilde\rho^2\, 
d\tau^2\right\} \nonumber\\
&&+\, 
 (\rho_+^2-b^2 \cos^2\theta) d\theta^2  
+ {\sin^2\theta \, (\rho_+^2-b^2)^2 \over  \rho_+^2-b^2\cos^2\theta} d\tilde \phi^2
\Bigg]\, ,
\een
\be\label{escaledgauge}
A_\mu dx^\mu \simeq i\, a\, {q \, \rho_+\over \rho_+^2 - b^2} \left[
d\tau-b\sin^2\theta \, {\rho_+^2 - b^2 \over \rho_+^2 - b^2 \cos^2\theta} d\tilde\phi
\right] - i\, a\, {q \, \rho_+\over \rho_+^2 - b^2}  \, d\tau\, .
\ee
We have dropped the   terms of order $\tilde\rho^2 d\tau d\tilde\phi$
and $\tilde\rho^2 d\tilde\phi^2$ since we shall argue shortly that
they represent non-singular terms in the metric
near the horizon. 
We now see that the $\tilde\rho$-$\tau$ space describes a smooth disk if we treat 
$\tilde\tau\equiv \tau\sqrt{m^2-q^2+b^2}/(\rho_+^2-b^2)$ as an angular coordinate with period $2\pi$,
since in the coordinate system $\tilde x=\tilde\rho\cos\tilde\tau$, $\tilde y = \tilde\rho\sin
\tilde\tau$ the metric is proportional to $d\tilde x^2 + d\tilde y^2$. Also in this 
coordinate system the $\tilde\rho^2 d\tilde\phi^2$ and $\tilde\rho^2 d\tilde\rho d\tilde\phi$
terms vanish at $\tilde x=\tilde y=0$.
This shows that in order to get a non-singular metric near the horizon, we need to
make the identification:
\ben
&& 
(\tau,\tilde\phi)\equiv \left(\tau + 2\pi \, {\rho_+^2-b^2\over \sqrt{m^2-q^2+b^2}}, \tilde\phi\right)
\nonumber \\
&\Rightarrow& 
(\tau,\phi)\equiv \left(\tau + 2\pi \,  {\rho_+^2-b^2\over \sqrt{m^2-q^2+b^2}}, \phi-
2\pi\, {b\over \sqrt{m^2-q^2+b^2}}\right)\, .
\een
This can be expressed as,
\be\label{egenperiod}
(\tau,\phi)\equiv  (\tau+\gamma, \phi - i\beta\, \Omega_3)\, .
\ee
The constant additive term in $A_\mu dx^\mu$ ensures that the integral of $A_\mu dx^\mu$
along the contractible cycle parametrized by $\tau$ vanishes at the horizon.

As reviewed in the introduction, the saddle point that contributes to the index has
$\beta\, \Omega_3=-2\pi i$. From \refb{esmall1} and \refb{esmall2} we see that
this can be achieved by choosing,
\be\label{emq}
m=q\, .
\ee
This gives
\be\label{eextremal}
\rho_\pm=q\pm b, \qquad \gamma= 2\pi {q (q+2b)\over b}, \qquad 
\gamma\omega_3=\beta\, \Omega_3 = -2\pi i,
\qquad \mu = {q+b\over q+2b}\, .
\ee

We can take the large charge limit by taking $a$ to be large at fixed $m(=q)$ and $b$.
With the choice of variables made in \refb{echoice}, the metric given in
\refb{escaledmetric} has an overall factor of
$a^2$ and the gauge field given in \refb{escaledgauge} has an overall factor of $a$. 
This makes it easy to extract the $a$ dependence of various quantities. In this limit
the curvature invariants at the horizon remain 
small so that the contribution from the
higher derivative corrections to the action are suppressed. As a result, the 
leading contribution to the entropy
is given by the Bekenstein-Hawking formula and the logarithmic corrections are the
dominant corrections to this formula.
We also
see that since $b$ and $q$ are free parameters, we can work at an arbitrary 
rescaled temperature
$\gamma^{-1}$ and chemical potential $\mu$ 
by adjusting $b$ and $q$. The physical temperature $\beta^{-1}$ scale as
$a^{-1}$  in the large $a$ limit.

\subsection{Integration measure over the zero modes} \label{sc}

In this section we shall review the $a$ dependence of the integration measure over
the various zero modes (and non-zero modes)
following \cite{1005.3044,1106.0080,1109.3706,1205.0971}.

First consider the case of a vector field $A_\mu$. Let us denote by $f^{(n)}_\mu$ the
basis functions for $A_\mu$, normalized so that $f^{(n)}_\mu$ does not have any
explicit $a$ dependence. Then we can expand $A_\mu$ as $\sum_n c_n f^{(n)}_\mu$.
The integration measure $\DD A$ over the modes $c_n$ are fixed so
that\footnote{Equivalently, we could keep the normalization of the zero modes arbitrary but
divide the path integral by $\int \DD A \, e^{-\int d^4 x \sqrt{ g} \, g^{\mu\nu} A_\mu A_\nu}$.}
\be\label{em1}
\int \DD A \, e^{-\int d^4 x \sqrt{ g} \, g^{\mu\nu} A_\mu A_\nu} =1\, .
\ee
Using the fact that $g_{\mu\nu}$ carries an overall factor of $a^2$ and $f^{(n)}_\mu$ is
$a$ independent, we see that the terms in the 
exponent scale as $a^2 c_m c_n$. Therefore \refb{em1}
requires that the integration measure over the modes $c_n$ must be chosen as
\be\label{em11}
\DD A \sim \prod_n d (a\, c_n)\, .
\ee

Next consider the case of metric deformation $h_{\mu\nu}$. 
Let us denote by $f^{(n)}_{\mu\nu}$ the
basis functions for $h_{\mu\nu}$, normalized so that $f^{(n)}_{\mu\nu}$ does not have any
explicit $a$ dependence. Then we can expand $h_{\mu\nu}$ as $\sum_n h_n f^{(n)}_{\mu\nu}$.
The integration measure $\DD h$ over the modes $h_n$ are fixed so that
\be\label{em2}
\int \DD h \, e^{-\int d^4 x \sqrt{ g} \, g^{\mu\nu} g^{\rho\sigma} h_{\mu\rho} h_{\nu\sigma}} =1\, .
\ee
Using the fact that $g_{\mu\nu}$ carries an overall factor of $a^2$ and $f^{(n)}_{\mu\nu}$ is
$a$ independent, we see that the terms in the exponent scale 
as $ h_m h_n$. Therefore \refb{em2}
requires that the integration measure over the modes $h_n$ must be chosen as
\be\label{em21}
\DD h\sim \prod_n d (h_n)\, .
\ee

Finally, let us consider the case of gravitino modes. This analysis is similar to that for the gauge
field. If $g^{(n)}_\mu$ is the $a$ independent basis for expanding the gravitino field $\psi_\mu$
and if we expand $\psi_\mu$ as $\sum_n \psi_n g^{(n)}_\mu$, then the integration measure 
$\DD\psi$ should satisfy
\be\label{em3}
\int \DD \psi \, e^{-\int d^4 x \sqrt{ g} \, g^{\mu\nu} \bar\psi_\mu \psi_\nu} =1\, .
\ee
Since the terms in the exponent scale as $a^2 \psi_n^\dagger \psi_m$, we get
\be\label{em31}
\DD \psi\sim \prod_n d (a\, \psi_n)\, .
\ee

\sectiono{Rotational zero modes} \label{sb}

Since the Kerr-Newman solution described in section \ref{sa} carries angular momentum
along the $z$-axis, we can generate zero modes via infinitesimal rotations about the $x$ and
$y$ axes. In this section we shall analyze the zero mode generated by rotation about the
$x$-axis. A similar analysis can be carried out for the zero mode associated with rotation about
the $y$-axis.

In the 
asymptotic region and Cartesian coordinates, the rotation by $\alpha_x$ along the $x$-axis
takes the form
\be
y\to \cos\alpha_x \, y + \sin\alpha_x \, z, \qquad z\to \cos\alpha_x\, z - \sin\alpha_x\, y\, .
\ee
In polar coordinates this translates to
\be\label{epolar}
\cos\theta \to \cos\alpha_x\cos\theta - \sin\alpha_x\sin\theta\sin\phi,
\qquad
\sin\theta\sin\phi \to \cos\alpha_x \sin\theta\sin\phi + \sin\alpha_x\, \cos\theta\, .
\ee
From this it is clear that the range of $\alpha_x$ is 0 to $2\pi$, independent of
the
overall scale $a^2$ that appears in the expression for the metric.  If we replace $\alpha_x$
by an arbitrary smooth 
function of the radial coordinate $\rho$ (and possibly other coordinates) such
that it approaches $\alpha_x$ as $\rho\to \infty$, then it will generate the same physical
configuration as that for constant $\alpha_x$
since the two differ by a general coordinate transformation localized in a
finite region of space-time. While applying this transformation to the Kerr-Newmann solution,
we shall choose the transformation parameter
such that it has the form:
\ben
\alpha_x && \hbox{for $\rho>\rho_0$,}\nonumber \\
0 &&  \hbox{for $\rho<\rho_1$ where $\rho_1<\rho_0$,} 
 \een
 and smoothly interpolates between $0$ and $\alpha_x$ in the range 
 $\rho_1<\rho<\rho_0$. Different choices of $\rho_0$ and $\rho_1$ are related
 by local general coordinate transformations and hence describe gauge equivalent
 solutions.
We
apply this
transformation on the Kerr-Newman solution to generate zero modes.
In this case since the
deformations vanish at the horizon, we do not need to worry about the regularity of the
modes at the horizon. On the other hand, since these transformations change the
physical angular momentum of the solution as long as $\alpha_x\ne 0$, and since the
angular momentum can be expressed as a surface integral at infinity, there will
be no general coordinate transformation localized in a finite region of space-time 
that can reduce the deformed configuration
to the original configuration. This ensures that these deformations are genuine deformations
and not just gauge artifacts. In the following, the formula for the deformations should be
understood as being valid for $\rho>\rho_0$.

We now consider the infinitesimal version of the transformation \refb{epolar}. 
This takes the form:
\ben
    \delta \q &=& \alpha_x \sin\phi, \\
    \delta \phi &=& \alpha_x \cot\q \cos\phi  ~,
\een
and hence
\ben
d(\delta \q) &= & \alpha_x \cos\phi \, d\phi, \\
d(\delta \phi) &=& -\alpha_x \Big[ \csc^2\theta \cos\phi\, d\q + \cot\q \sin\phi\, d\phi \Big]\, .
\een
Using these we can compute the value of $\delta(ds^2)$ and $\delta A_\mu$
for the Kerr-Newman metric.
These are given as 
\ben\label{erotmetric}
\delta(ds^2)&=& a^2 \alpha_x \Bigg[  -\frac{b^2   \sin (2 \theta) \, \sin \phi  \left(q^2-2 m \rho\right)}{(\rho^2 - b^2 \cos^2 \theta)^2} d\tau^2+    \frac{ b^2  \sin (2\theta)  \, \sin \phi }{\rho^2-b^2
-2 m \rho+q^2} d\rho^2 \nonumber\\
&&    +\frac{b   \cos \phi  \left(q^2-2 m \rho\right)}{(\rho^2 - b^2 \cos^2 \theta)} d\tau d\q  
-  \frac{ b   \sin (2 \theta) \, \sin \phi  \left(q^2-2 m \rho\right) \left(\rho^2 - b^2 - b^2 \sin^2\q\right)}{(\rho^2 - b^2 \cos^2\theta)^2} d\tau d\phi 
\nonumber\\ &&   
+b^2   \sin (2\theta) \,  \sin \phi \, d\q^2 
+\frac{2 b^2 \sin ^2\theta \, \cos \phi \, \left(\rho^2 - b^2 \cos^2 \q + 2m \rho
- q^2\right)}{(\rho^2 - b^2 \cos^2\theta)} d\q d\phi \nonumber\\ &&  
+\frac{2 b^2   \left(\rho^2-b^2\right) \sin ^3\theta \, \cos \theta \, \sin \phi \, \left(q^2-2 m \rho\right)}{(\rho^2 - b^2 \cos^2\theta)^2} d\phi^2 \Bigg]\, ,
\een
and
\ben\label{erotgauge}
\delta(A_\mu dx^\mu) &=& i \, a\, \alpha_x\, q \, b\, \rho\Bigg[ 
-\frac{2   b  \sin \theta  \cos \theta  \sin \phi }{\left(\rho ^2-b^2 \cos ^2\theta \right)^2} d\tau +\frac{   \cos \phi }{\rho ^2-b^2 \cos ^2\theta }\, d\q \nonumber\\ &&\hskip 1in -\, 
\frac{  \sin 2 \theta  \sin \phi  \left(\rho^2 - b^2 - b^2\sin^2\theta \right)}{2
\left(\rho ^2-b^2 \cos ^2\theta \right)^2} \, d\phi\Bigg]~.
\een
These have the asymptotic fall-off $\delta h_{\mu\nu}\sim \rho^{-2}$, $\delta A_\mu\sim\rho^{-2}$
in Cartesian coordinates, and
hence describe normalizable modes. Furthermore, while for generic Kerr-Newman solution they
would not satisfy the periodicity condition \refb{egenperiod},
for the  $\beta\, \Omega_3=-2\pi i$ solution described in \refb{emq} and \refb{eextremal},
\refb{erotmetric}  and \refb{erotgauge}
certainly satisfy the required periodicity under $(\tau,\phi)\to
(\tau+\gamma,\phi-2\pi)$. Therefore this deformation describes
genuine zero mode. A similar analysis can be used to show that rotation about the $y$-axis
with parameter $\alpha_y$
also generates a normalizable zero mode.

Finally, note that $a^2\alpha_x$ is an overall factor in \refb{erotmetric}. Therefore when we change
variables from integration over the modes $h_n$ of $h_{\mu\nu}$ to the parameters $\alpha_x$,
$\alpha_y$,
we pick up a jacobian
factor of $a^2$ for each of these modes. The same conclusion is reached if we analyze the
modes $c_n$ of the gauge fields, since according to \refb{em11} 
the integration measure over the
modes $c_n$ is $a dc_n$ and we have an overall factor of $a$ in \refb{erotgauge}. The actual
zero mode is proportional to 
a linear combination of the modes $h_n$ and $a\, c_n$ as given by 
\refb{erotmetric}, \refb{erotgauge}.

\section{Logarithmic corrections to the entropy} \label{slog}
 
 Our goal in this section will be to compute the logarithmic correction to the entropy of
 supersymmetric black holes in theories with $\NN\ge 2$ supersymmetry.
As was argued in \cite{1108.3842}, the logarithmic correction to the black hole entropy is
not sensitive to the details of the prepotential of the theory 
and so we can compute this by taking a simple
prepotential proportional to $(X^0)^2 - \sum_{i=1}^{n_V} (X^i)^2$. Furthermore
we can assume, without loss of generality, that the black hole
carries  charge of only the 0-th U(1) gauge field that we identify 
as the graviphoton of the $\NN=2$ supersymmetric theory. In this case we can 
construct a black hole solution for which the vector multiplet
scalars $X^i/X^0$ remain zero and the solution takes the form of a Kerr-Newman solution.
Furthermore, using the electric magnetic duality transformation, we can work with
a black hole carrying only electric charge and no magnetic charge. 
If $Q$ denotes this electric charge
then BPS saturation gives $M=Q$.

For zero temperature, zero angular momentum 
geometry, the logarithmic correction to the entropy of such black holes
was computed in \cite{1108.3842,1404.1379} by evaluating the integrals over the 
fluctuations of all the massless fields in the near horizon 
$AdS_2\times S^2$ background geometry. Our goal will be to compute this using the
finite temperature geometry and then compare the result with that obtained using the
zero temperature geometry. For finite temperature, the 
relevant geometry is the full
Kerr-Newman geometry with the angular velocity of the black hole 
set to $(0,0,-2\pi i)$. As discussed in section
\ref{sa}, the inverse
temperature $\beta$ is arbitrary.
We shall take 
$\beta \sim M$. In this case we can work in a coordinate system such that 
under an overall scaling of $Q$ by $a$, the metric has an overall scale factor
$a^2$ and the area $A$ of the event horizon scales as $a^2$.  Explicit form of this
metric has been given in \refb{eknscaled}.
We shall be interested in computing corrections to the black hole
entropy of order $\ln a^2\sim \ln A$,

There is one point that needs to be kept in mind in carrying out this comparison. The
computation in the zero temperature 
near horizon geometry gives the result in the microcanonical ensemble
while the computation in the finite temperature full Kerr-Newman geometry gives the
result in the grand canonical ensemble. Therefore {\it a priori} we need to make a 
Laplace transform before we can compare the two results, since the Laplace transform
can generate additional logarithmic terms. However,
one can see as follows that 
in $D=4$ we do not need any correction from the change of ensemble.
Since the entropy scales as $a^2$ under the scaling of charges by $a$ in $D=4$, the
second derivative of the entropy with  respect to the charges does not scale with $a$.
Therefore no new logarithmic corrections are generated as we go between the
fixed charge and fixed chemical potential system.
This can also be seen from eq.(3.12) of \cite{1205.0971}. 
From the finite temperature perspective, since
we want to compute the index, we must
sum over all angular momentum states and hence as far as the angular momentum is 
concerned, we do need to work in the grand canonical ensemble. 
In contrast in the zero temperature
near horizon geometry 
the angular momentum is fixed to zero by requirement of supersymmetry.
The fermion zero modes associated with broken supersymmetry, that are needed to
form the supermultiplet carrying different spins, live outside the
horizon and produce at most a finite multiplicative factor in the expression for the index. 
Hence up to this factor,
the entire
contribution to the index can be expressed in terms of the contribution 
from zero angular momentum states and this should agree with the result for the index
computed using the finite temperature geometry (up to factors of order unity coming from the
change of ensemble).

\subsection{Non-zero mode contributions}

The logarithmic
terms come from two sources: from path integral over
the non-zero modes and path integral over the zero modes. Since the kinetic operator
for the bosons has eigenvalues of order $a^{-2}$ and the kinetic operator for the fermions
has eigenvalues of order $a^{-1}$, the integration over the zero modes will produce
a net contribution of order $\left(N_B-{1\over 2} N_F\right)\ln a={1\over 2}
\left(N_B-{1\over 2} N_F\right)\ln A$ to the entropy, where $N_B$ and $N_F$ are the total
number of bosonic and fermionic non-zero modes respectively. This can be expressed
as\cite{duffobs,christ-duff1,christ-duff2,duffnieu,birrel,gilkey,0306138}:
\be\label{e2}
{1\over 2} \, \ln A\, \int d^4 x \, a_4(x)\, ,
\ee
where $a_4(x)$ is the fourth Seeley - DeWitt coefficient\cite{sd,dewitt,0306138}
of the massless fields in the theory.\footnote{$\int d^4x  \, a_4(x)$  actually computes
$N_B-{1\over 2} N_F$ where $N_B$ and $N_F$ counts all bosonic and fermionic
modes, including the
zero modes. As a result, \refb{e2} also includes a contribution
from the zero modes, treating them as if they have non-zero eigenvalues of the kinetic
operator that scale in the same way as the eigenvalues of the non-zero modes. These 
contributions have to be removed in the final result. \label{f4}
}
For minimally coupled charge neutral fields the heat kernel is determined in terms of 
background Riemann tensor. An example of this is the logarithmic contribution to the
entropy due to a minimally coupled scalar analyzed in \cite{9709064}.  
However, for fields in supergravity that couple to both the background
metric and gauge fields in a non-minimal way, $a_4$
in general depends on the background Riemann tensor and gauge field strengths.
A surprising fact, observed in \cite{1505.01156} and later verified in
\cite{1905.13058,2012.12227}, 
is that for $\NN\ge 2$ supergravity, for which we can 
divide the massless fields as belonging to $n_H$ hypermultiplets, $n_V$ vector
multiplets, one  gravity multiplet and $(\NN-2)$ gravitino multiplet, $a_4$ is 
proportional to the Euler density:
\be\label{e3}
a_4=-{1\over 16\times 24\pi^2} \, [-11 + 11(\NN-2)+n_V-n_H] \, E_4, 
\qquad E_4 = R_{\mu\nu\rho\sigma} R^{\mu\nu\rho\sigma}-4 R_{\mu\nu} R^{\mu\nu} 
+R^2\, .
\ee
In particular, the dependence on the background gauge fields can be expressed in terms
of the Ricci tensor after using Einstein's equation and the equations of motion and the
Bianchi identities of the gauge field strengths.
Since the Euler density is a topological term, its integral over the near horizon 
$AdS_2\times S^2$ geometry and the full geometry gives the same result (after adding
appropriate boundary terms for $AdS_2\times S^2$). This has been explicitly
verified in appendix \ref{curvature}.  In both cases $\int d^4x \sqrt g\, E_4=64\pi^2$ and
\refb{e2} gives
\be\label{e4}
- {1\over 12} \, [-11 + 11(\NN-2)+n_V-n_H] \,\ln A \, .
\ee
Since \refb{e4} is the same for the zero temperature and finite temperature 
geometry, we only need to check if
the zero mode contributions agree between the two computations.

\subsection{Zero mode contribution}

Zero modes are generated by symmetry transformations of the asymptotic geometry that are
broken by the solution. The transformations parameters are non-normalizable but the
deformations they produce are normalizable.
We shall compare the zero mode contributions from each super-multiplet
separately, since our goal
is to establish that the agreement between the calculations at zero temperature and finite
temperature agree for an arbitrary $\NN\ge 2$ supergravity.

\subsubsection{Hypermultiplet zero modes}

The hypermultiplet fields, consisting of scalars and spin half fermions,
do not have any zero modes either in the zero temperature $AdS_2\times S^2$
geometry or in the finite temperature geometry. Therefore
their contribution to the logarithmic corrections agree trivially between zero
temperature computation and finite temperature computation.

\subsubsection{Vector multiplet zero modes}

The gauge fields
of the vector multiplet have zero modes in the zero temperature geometry, 
but the contribution from the zero modes turns out to be the same that would have been
there if these modes had been non-zero modes\cite{1108.3842}. 
Therefore the heat kernel already captures
their contribution and no further corrections are necessary (see footnote \ref{f4}).
For the finite temperature
geometry there are no zero modes for the vector multiplet fields\cite{1205.0971}. 
Therefore the contribution from the vector multiplet zero modes also agree 
trivially between the zero and finite
temperature calculations. 

\subsubsection{Gravity multiplet zero modes}

Next we turn to the gravity multiplet zero modes. In the zero temperature geometry there
are both fermionic and bosonic zero modes. In particular 
we know from eq.(5.31) of \cite{1108.3842} 
that the net logarithmic contribution to the
entropy from the gravitino zero modes is given by $4\ln A$ and from
eqs.(2.19) and (4.39) of 
\cite{1108.3842} 
that the net logarithmic correction due to the metric and graviphoton zero modes is given
by $-3\ln A$.  These factors include the terms that need to be subtracted from the
heat kernel contribution due to the fact that the heat kernel includes contributions from the
zero modes treating them as non-zero modes.
Therefore the net logarithmic corrections due to the zero modes in the zero temperature
geometry is
\be\label{e14}
\ln A\, .
\ee

We shall now carry out the analysis in the finite temperature geometry. 
Eq.(2.41) of \cite{1205.0971} 
tells us that in $D=4$ there is no logarithmic correction from the
zero modes of the metric. 
There is, however, one subtle point we need to take into account. A priori a rotating
black hole in the Lorentzian geometry 
carries three translational zero modes generated by translation along
the three spatial directions and two rotational zero modes corresponding to  rotations about
the two axes perpendicular to the axis of rotation. However in the Euclidean geometry, corresponding to a  black hole rotating around the third axis
with angular velocity $\Omega_3$ at the horizon, we require the fields to be periodic under simultaneous
translation of the Euclidean time $\tau$ 
by $\beta$ and the azimuthal angle $\phi$ by $-i\beta\, \Omega_3$.
For generic $\beta\, \Omega_3$,  it was argued in \cite{1205.0971}
that only the translational zero mode along the $z$ direction satisfies this
periodicity requirement. The other two
translational modes and the rotational modes have non-trivial dependence on
$\phi$ and fails to satisfy the periodicity requirement. However, in the present
context we have $\beta\, \Omega_3=-2\pi i$ and so the modes are required to be
periodic under $\tau\to\tau+\beta$, $\phi\to \phi-2\pi$. Since the modes are
$\tau$ independent, this only demands that they are periodic under $2\pi$ shift of
$\phi$ and this is automatically satisfied.  
This has been illustrated in 
section \ref{sb} for the rotational zero modes.
This means that we now need to take into account 
the contribution from two extra translation zero modes and two extra rotational zero
modes.

Now it was shown in \cite{1205.0971} that in $D$ space-time dimensions the contributions 
from
the translational zero modes are proportional to $(D-4)$.
Therefore they do not contribute for $D=4$.\footnote{If we construct the
translation zero modes using a procedure similar to that in section \ref{sb}, then, due to the
overall factor of $a^2$ present in the metric, the range of each
transformation parameter is of order $L/a$ where $L$ is the proper length of a box in which
we confine the system. This is to be contrasted with the parameters of rotational zero modes
which have $a$ independent range. Therefore integration over each translation zero mode
will produce an extra factor of $1/a$ compared to that over a rotational zero mode.
This translates to an extra additive factor of $(-\ln a)$ 
to the entropy for each translation zero
mode and is the
main difference between
the contributions from translational zero modes and the rotational zero modes.
}
 This leaves us with the two rotational zero
modes described in section \ref{sb}. 
In order to evaluate their contribution we use \refb{em21}  to
conclude that if we expand the metric fluctuation $h_{\mu\nu}$ as  
$\sum_n h_n f^{(n)}_{\mu\nu}$ where
$f^{(n)}_{\mu\nu}$ are basis functions, normalized so that they do not carry any $a$ dependence,
then the integration measure over the modes $h_n$ should come as 
\be
\prod_n dh_n\, .
\ee
In particular there is no $a$ dependence of the measure. For the zero modes,
we can now change variable of integration from  $h_n$ to the parameters $\alpha_x,\alpha_y$
describing rotational zero modes.
As described in the last paragraph of 
section \ref{sb}, for each zero mode, this introduces a jacobian factor of $a^2$.
Since the $\alpha_x,\alpha_y$ integration ranges are $a$ independent, it produces a net
factor
of $(a^2)^2$.
On the other hand, the heat kernel result counts the contribution
from each mode (including the zero modes) as $a$, since the non-vanishing eigenvalues of the
kinetic operator are of order $1/a^2$. Therefore we need to divide the partition function by
a factor of $a$ for each zero
mode $h_n$. This yields a net contribution of
\be\label{e7}
\ln {(a^2)^2\over a^2}= 2\, \ln a \sim \ln A\, ,
\ee
to the entropy.

We now turn to the contribution due to the
gravitino zero modes of the gravity multiplet.
As described in eq.\refb{em31}, 
if we expand the gravitino field $\psi_\mu$ as $\sum_n \psi_n g^{(n)}_\mu$ where
$g^{(n)}_\mu$ are basis functions, normalized so that they do not carry any $a$ dependence,
then the measure over the gravitino modes should come as 
\be\label{e15}
\prod_n d(a \psi_n)\, .
\ee 
Naively integral over these gravitino zero modes will vanish
but this is avoided as follows. As mentioned in the introduction, for every pair of
broken supersymmetry, we need to insert a factor of $2J_3$ in the path
integral. 
Since a supersymmetric black hole has four unbroken supersymmetries and
since $\NN=2$ supergravity has eight unbroken supersymmetries in Minkowski space,
the black hole breaks four of these eight supersymmetries. 
So we have a net factor of $(2J_3)^2$. We shall now show that the presence of this factor
of $(2J_3)^2$ makes the zero mode integral non-vanishing.

We shall construct $J_3$ using the Noether procedure. First note that 
the action has the form
\be\label{e2.7}
\int d^4 x \, \sqrt{ g} \,  \bar \psi_\mu \gamma^{abc} \p_\rho \, \psi_\nu \, E_a^\mu E_b^\nu
E_c^\rho \sim a\, \int d^4 x \bar \psi \p \psi \, 
\ee
where $E^\rho_a$ is the inverse vierbein. 
Since $J_3$ is the generator of rotation, it acts on the field $\psi$ without
any factor of $a$. Therefore the contribution to $J_3$ from the fermions, obtained using
the Noether procedure from the action \refb{e2.7}, is a linear combination of terms of the form
$a\, \psi_m \psi_n$. 
If we denote by $\psi_1,\cdots, \psi_4$ the four fermion zero modes then 
that the relevant part of $(2J_3)^2$
is $a^2 \psi_1\cdots \psi_4$. Using \refb{e15} we now see that the integration over the
zero modes give
\be\label{e417}
\prod_{n=1}^4 d(a\psi_n) \, a^2 \psi_1\cdots \psi_4 \sim a^{-2}\, .
\ee

We need to divide this by the power of $a$ that is generated
by the contribution of the zero modes
to the
heat kernel since the heat kernel treats the zero modes as if they were non-zero modes.
Indeed, the expression based on the heat kernel 
counts a factor of $a^{-1}$ for each pair of gravitino zero modes, since this is how
the non-zero
eigenvalues of the Dirac operator scales. Therefore the heat kernel contribution includes
a net factor of $a^{-2}$ from the
four gravitino zero modes which need to be removed from the expression for heat kernel
contribution to the index. Dividing \refb{e417} by $a^{-2}$, 
we see that the net excess contribution due to
the gravitino zero modes, that is not counted in the heat kernel analysis, 
is independent of $a$.
Therefore it gives
zero net logarithmic correction to the entropy. 

Combining this result with \refb{e7} we get the net logarithmic
contribution to the index from the gravity multiplet zero modes to be
\be\label{e8}
\ln A\, .
\ee
This agrees with the result \refb{e14} of the zero temperature analysis.

\subsubsection{Gravitino multiplet zero modes}

Finally we turn to the extra gravitino
multiplets that are present in $\NN>2$ supergravity theories. 
These have no zero modes in the zero temperature sector, as can be seen from
the analysis of the zero mode structure for black holes in $\NN=4$ and $\NN=8$
supergravity. In the finite
temperature geometry they lead to $4(\NN-2)$ fermion zero modes that have the
following origin. Since a supersymmetric black hole preserves four supersymmetries
and an $\NN\ge 2$ supergravity has $4\NN$ unbroken supersymmetries in 
Minkowski space, $4(\NN-1)$
supersymmetries are broken by the black hole background. This leads to $4(\NN-1)$
goldstino fermion zero modes. Of these four fermion zero modes have already been 
attributed
to the zero modes of the gravitinos belonging to the gravity multiplet
of $\NN=2$ supergravity. The remaining $4(\NN-2)$
zero modes must come from
the $(\NN-2)$ extra gravitino multiplets.  The analysis of logarithmic contribution due to
these zero modes follows an analysis identical to that for the gravitino zero modes
coming from the gravity multiplet, and we are led to the same conclusion that they do not
give any additional logarithmic contribution besides those already counted in the heat
kernel.

This shows that the contribution to the logarithmic correction in the zero temperature and
finite temperature backgrounds agree for each super-multiplet separately.

\sectiono{Conclusion} \label{sconc}

In this paper we have used the formalism developed in \cite{2107.09062} to compute logarithmic
corrections to the logarithm of supersymmetric black hole index by working with a finite
temperature solution. The result is in perfect agreement with the zero temperature, near
horizon computation of the same quantity. 

Since the two procedures are different, the equality of the final results gives confirmation of both
methods. This also gives additional confirmation of the final results for the logarithmic 
corrections, many of which have been
tested in string theory via microscopic counting. Furthermore, since the finite temperature method
for computing the index is closely related to the computation of the logarithmic correction to the
entropy of non-supersymmetric black holes, the result of this paper also gives confidence in the
results for non-supersymmetric black holes, for which there is as yet
no independent verification of the
results from microscopic counting of states.

\bigskip

\noindent{\bf Acknowledgement}: 
We thank Alok Laddha for useful discussions.
Research at ICTS-TIFR is supported by the Department of Atomic Energy Government of India, under Project Identification No. RTI4001.
The work of A.S. is supported by ICTS-Infosys Madhava 
Chair Professorship
and the J.~C.~Bose fellowship of the Department of Science and Technology.

\begin{appendix}
\sectiono{Comparing the heat kernel contributions in the zero and finite temperature
computations} \label{curvature}

In this appendix, we describe explicit checks on the equality between the heat kernel 
contributions
in the zero temperature computation and the finite temperature computation.

In the usual literature, the metric for the Kerr-Newman solution 
is expressed in terms of Boyer-Lindquist coordinates described in \eqref{knboyer}. For our purpose, we find that it is easy to compute the curvature invariants like $R_{\mu\nu} R^{\mu\nu}$, $R_{\mu\nu\rho\sigma} R^{\mu\nu\rho\sigma}$, $F_{\mu\nu} F^{\mu\nu}$, etc. by going to the null-Kerr coordinates which are defined by the following coordinate transformations 
\be
du = dt - \frac{r^2 + a_0^2}{\Delta_0} dr~, \qquad 
d\tilde \phi = d\phi - \frac{a_0}{\Delta_0}  dr~,
\ee
where $\Delta_0 = r^2 + a_0^2 - 2 M r + Q^2$. In these coordinates, the metric \eqref{knboyer} becomes
\be
    g_{\mu\nu} =\left( 
\begin{array}{cccc}
    - 1 + \frac{2M r - Q^2}{r^2 + a_0^2 \cos^2 \theta} & -1 & 0 & \frac{a_0 \sin^2 \theta}{r^2 + a_0^2 \cos^2 \theta} (Q^2 - 2 M r)  \\
     -1 & 0 & 0 & a_0 \sin^2 \theta \\
     0 & 0 & r^2 + a_0^2 \cos^2 \theta & 0 \\
     \frac{a_0 \sin^2 \theta}{r^2 + a_0^2 \cos^2 \theta} (Q^2 - 2 M r)  & a_0 \sin^2\theta & 0 & \frac{\sin^2 \theta}{r^2 + a_0^2 \cos^2 \theta} 
     \big((r^2 + a_0^2)^2-\Delta_0 a_0^2 \sin^2 \theta \big)
\end{array}
    \right)~.
\ee
The inverse metric is
\be
    g^{\mu\nu} = \frac{1}{r^2 + a_0^2 \cos^2 \theta} \left( 
\begin{array}{cccc}
     a_0^2 \sin^2 \q & -(r^2 + a_0^2) & 0 & - a_0  \\
      -(r^2 + a_0^2) & \Delta_0 & 0 & -a_0 \\
     0 & 0 & 1 & 0 \\
     -a_0 & -a_0 & 0 &  \frac{1}{\sin^2 \theta}
\end{array}
    \right)~.
\ee
This form  of $g^{\mu\nu}$ makes the computation of the curvature invariants easy. It is now simple to check that we recover the result of $R^{\mu\nu} R_{\mu\nu}$ and $R^{\mu\nu\rho\sigma}R_{\mu\nu\rho\sigma}$ for Kerr-Newman geometry as given in \cite{9912320,0302095,Bhattacharyya:2012wz},
\ben\label{R2}
R_{\mu\nu} R^{\mu\nu} &=& \frac{4 Q^4}{(a_0^2 \cos^2 \theta + r^2)^4}~,\\
R_{\mu\nu\rho\sigma} R^{\mu\nu\rho\sigma} &=& \frac{8}{\left(a_0^2 \cos^2\theta +r^2\right)^6} \Big[  -12 M Q^2 r \left(5 a_0^4 \cos^4\theta-10 a_0^2 r^2 \cos ^2\theta+r^4\right)\nonumber\\
&&+Q^4 \left(7 a_0^4 \cos^4\theta -34 a_0^2 r^2 \cos^2\theta+7 r^4\right) \label{R2a}\\
&&+6 M^2 \left(-a_0^6 \cos^6\theta+15 a_0^4 r^2 \cos ^4\theta-15 a_0^2 r^4 \cos^2\theta+r^6\right)\Big]\, . \nonumber
\een
We can also compute other curvature invariants and obtain compact expressions, for example,
\be
F^{\mu\nu} F_{\mu\nu} = \frac{4 Q^4 \left(a_0^4 \cos ^4\theta -6 a_0^2 r^2 \cos ^2\theta +r^4\right)^2}{\left(a_0^2 \cos ^2\theta +r^2\right)^8} \, ,
\ee
\ben
    R_{\mu\nu\rho\sigma} F^{\mu\nu}F^{\rho \sigma} &=& \frac{4 Q^2 \left(14 a_0^6 M r \cos ^6\theta -70 a_0^4 M r^3 \cos^4\theta +42 a_0^2 M r^5 \cos ^2\theta -2 M r^7\right)}{\left(a_0^2 \cos^2\theta +r^2\right)^7}\nonumber\\
    &&\quad +\frac{4 Q^4 \left(-a_0^6 \cos^6\theta +33 a_0^4 r^2 \cos^4\theta -27 a_0^2 r^4 \cos^2\theta +3 r^6\right)}{\left(a_0^2 \cos^2\theta +r^2\right)^7}~.
\een
We now evaluate $\int d^4 x\ a_4$  using equations \eqref{R2} and \refb{R2a}.
Since $a_4$ given in \refb{e3} is proportional to the Euler density $E_4$ \cite{1505.01156}, 
it suffices to consider the integral of $E_4$.  For the geometries we are studying 
(where $R = 0$), $E_4$ 
is proportional to $R_{\mu\nu\rho\sigma} R^{\mu\nu\rho\sigma} - 4 R_{\mu\nu} R^{\mu\nu}$. By using equations \eqref{R2}, \refb{R2a} and \eqref{beta}, we obtain the following result:
\be\label{KNR2}
 \int d^4 x \sqrt{g}\ (R_{\mu\nu\rho\sigma}R^{\mu\nu\rho\sigma} - 4 R_{\mu\nu} R^{\mu\nu}) = 64 \pi^2~.
\ee

We now derive the 
corresponding 
result in the zero-temperature geometry. The metric of $AdS_2 \times S^2$ is given as 
\be
ds^2 =a^2 \left[ (r^2- 1) \, d\tau^2 + \frac{dr^2}{r^2 - 1}  + 
(d\theta^2 + \sin^2 \theta d\phi^2) \right]~.
\ee
The horizons are located at $r = \pm 1$. In this metric, the curvature invariants take the following form
\be
 R_{\mu\nu\rho\sigma}R^{\mu\nu\rho\sigma}= \frac{8}{a^4}, \qquad 
 R_{\mu\nu} R^{\mu\nu}= \frac{4}{a^4}~.
\ee
This gives us,
\be
E_4 = R_{\mu\nu\rho\sigma}R^{\mu\nu\rho\sigma} - 4 R_{\mu\nu} R^{\mu\nu}  
= - \frac{8}{a^4}~.
\ee
By integrating the expression above, we obtain,
\be
\int d^4 x \sqrt{g} \, E_4 = 64 \pi^2(1 - r_0)~,
\ee
where $r_0$ is the upper limit of the $r$-integral which we eventually take to infinity. 
The term proportional to $r_0$ is 
removed by adding an appropriate counterterm proportional to
the length of the boundary since this does not contribute to the ground state 
degeneracy\cite{0809.3304}. 
The finite part of the result, that contributes to the log of the degeneracy, 
is the same as in the Kerr-Newmann geometry given in \eqref{KNR2}.

\end{appendix}

\end{document}